\documentclass[11pt]{aastex}
\usepackage{graphicx,emulateapj5,apjfonts}
\usepackage{onecolfloat}
\usepackage{epsf}
\bibpunct {(} {)} {;} {a} {} {,}	

\shortauthors{Ruhland et al.}
\shorttitle{Evolution of Early-type galaxies}

\slugcomment{{\sc Astrophysical Journal in press: } January 27, 2009 }

\newcommand{\combo}{{\sc COMBO-17}}
\newcommand{\gems}{{\sc Gems}}

\newcommand{\goods}{{\sc Goods}}

\newcommand{\galfit}{{\sc Galfit}}

\newcommand{\pegase}{{\sc P\'egase }}

\begin{document}

\def\head{

\title{The evolution of the scatter of the cosmic average color-magnitude relation: Demonstrating consistency with the ongoing formation of elliptical galaxies}

\author{Christine Ruhland$^1$, Eric F.\ Bell$^1$, Boris H\"au\ss{}ler$^{1,2}$, Edward N.\ Taylor$^3$, Marco Barden$^{1,4}$, Daniel H.\ McIntosh$^{5,6}$}

\affil{$^1$ Max-Planck-Institut f\"ur Astronomie, K\"onigstuhl 17, D-69117 Heidelberg, Germany; \texttt{ruhland@mpia.de}\\ 
$^2$ University of Nottingham, Nottingham NG7 2RD United Kingdom \\ 
$^3$ Sterrewacht Leiden, Leiden University, P.O. Box 9513, NL-2300 RA Leiden \\
$^4$ Institut f\"ur Astro- und Teilchenphysik, Universit\"at Innsbruck, Technikerstra\ss{}e 25/B, A-6020 Innsbruck \\
$^5$ Astronomy Department, University of Massachusetts, Amherst, MA 01003 USA \\ 
$^6$ Department of Physics, University of Missouri - Kansas City, Kansas City, MO 64110 USA}

\begin{abstract}

We present first measurements of the evolution of the scatter of the cosmic average early-type galaxy color--magnitude relation (CMR) from $z=1$ to 
the present day, finding that it is consistent with models in which 
galaxies are constantly being added to the red sequence through truncation 
of star formation in blue cloud galaxies.  We used a sample of over 700 
red sequence, structurally-selected early-type galaxies (defined to have S\'ersic index $>2.5$) with redshifts 
$0<z<1$ taken from the Extended Chandra Deep Field South (173 galaxies) and the Sloan 
Digital Sky Survey (550 galaxies), constructing rest-frame $U-V$ colors accurate to 
$<0.04$\,mag. We find that the scatter of the CMR of cosmic average early-type 
galaxies is $\sim0.1$\,mag in rest-frame $U-V$ color at $0.05<z<0.75$, and 
somewhat higher at $z=1$.  We compared these observations with a model in 
which new red sequence galaxies are being constantly added at the rate 
required to match the observed number density evolution, and found that this 
model predicts the correct CMR scatter and its evolution.  Furthermore, 
this model predicts approximately the correct number density of `blue 
spheroids' --- structurally early-type galaxies with blue colors --- 
albeit with considerable model dependence.  Thus, we conclude that both 
the evolution of the number density and colors of the early-type galaxy 
population paint a consistent picture in which the early-type galaxy 
population grows significantly between $z=1$ and the present day through 
the quenching of star formation in blue cloud galaxies.

\end{abstract}

\keywords{galaxies: evolution --- galaxies: general ---
galaxies: elliptical and lenticular ---
galaxies: stellar content --- surveys }
}

\twocolumn[\head]

\section{Introduction}

One of the best-known and most powerful scaling relations 
of the early-type (elliptical and lenticular) galaxy population is 
the systematic reddening of their colors with increasing 
luminosity: the color--magnitude relation (CMR). The slope 
of the CMR is driven by a correlation between metallicity
and mass \citep{FaberJackson76,Kodama97,Terlevich99,Trager2000b,Gallazzi06}, while the scatter
is determined by scatter in both age and metallicity, where
it is generally thought that age is the dominant driver (\citealp{BLE92,Trager2000b,Gallazzi06}, although \citealp{Trager2000b} argue that an anticorrelation 
between age and metallicity keeps the scatter of the early-type
galaxy scaling relations relatively modest while allowing significant
scatter in both age and metallicity). 
The scatter in this correlation 
is relatively small \citep{Baum62,VisvanathanSandage1977,BLE92,Terlevich01,McIntosh2005cluster}.
Because the color of 
stellar populations is strongly affected by their ages and metallicities \citep{Worthey94}, 
this correlation is a powerful probe of the formation and evolution 
of the stellar populations in early-type galaxies. The small intrinsic scatter found 
in some clusters (as little as $\sigma_{U-V} = 0.04$\,mag in
 Virgo and Coma; \citealp{BLE92,Terlevich01}; although other clusters 
can have scatters approaching 0.1 mag; \citealp{McIntosh2005cluster}) 
gave considerable momentum to the notion that 
early-type galaxies formed the bulk of their stars at early times and that 
their stellar populations have aged essentially passively to the 
present day.

In the last few years, this position has been challenged by evidence
from deep redshift surveys that the cosmic average red sequence 
galaxy population (i.e., averaged over
all environments) builds up in stellar mass by roughly a factor of two 
over the interval $z=1$ to $z=0$ through the addition 
of new red sequence galaxies \citep[][although some of the papers argue for a build-up in stellar mass only in galaxies with $\la 10^{11} M_{\sun}$]{Chen,Bell04b,Cimatti06,Brown06,Faber07,Scarlata07}.  
These `new' red sequence galaxies are the result of truncation 
of star formation in some fraction of the blue cloud 
population \citep{Bell07} through, e.g., galaxy-galaxy merging 
\citep{Bell04b,Faber07,Hopkins07} 
or environmental processes such 
as strangulation or ram-pressure stripping \citep[e.g.,][]{Kodama2001}.

Such a scenario makes strong predictions about what the scatter of the CMR
and its evolution should be as a function of redshift \citep{vanDokkumFranx01}; to first order the scatter is expected to be constant with redshift. The object of 
this paper is to quantitatively test this picture. We carefully measure the CMR evolution of galaxies selected to have concentrated light profiles and red colors --- taken as a proxy for the early-type galaxy population --- from $z=1$ to the present day, and compare
it to a toy model of a growing red sequence.  We use the SDSS for 
the low-redshift CMR measurement, and a sample of galaxies
with spectroscopic redshifts and accurate HST colors from the 
extended Chandra Deep Field South (CDFS hereafter) to probe the CMR out to $z=1$. As it was {\it a priori} unclear how large the CMR scatter should have been, we adopted a conservative approach that optimized rest-frame color accuracy at the expense of sample size, choosing galaxies with spectroscopic redshifts and within relatively narrow redshift slices (to minimize $k$-correction uncertainties) and with color HST imaging (to minimize color measurement error). The data are described in \S 2.  We describe the 
$k$-corrections and their uncertainties for the intermediate redshift sample
in \S 3.  We present our measurements of the intercepts and 
scatter of the CMR for color and structurally selected samples as a function of redshift in \S 4.  In \S 5, 
we compare the observed results to stellar population models
of increasing complexity, finally comparing it to a model
for the growth of the red sequence through the truncation 
of star formation in blue cloud galaxies.  We present our conclusions 
in \S 6.  The casual reader may wish to skip to \S 5 directly.
In what follows, we assume $\Omega_{m,0} = 0.3$, $\Omega_{\Lambda,0} = 0.7$, 
and $H_0 = 70$\,km\,s$^{-1}$\,Mpc$^{-1}$ and a universally-applicable 
\citet{Kroupa01} stellar IMF for stellar population modeling.

\section{Data} \label{data}

We use data from two different sources, depending on redshift.  For 
galaxies with $0.5<z<1$, we choose to determine accurate colors for 
galaxies with spectroscopic redshifts in the HST/\gems{} survey of the CDFS.  
This is crucial for minimizing the color measurement error, placing the 
strongest possible constraints on the intrinsic scatter of the CMR.  To 
complement this dataset at low redshift, we use a sample drawn from the 
SDSS at $z=0.05$.

\subsection{Intermediate redshift data}
\begin{figure}[t]
	\centering
	\epsscale{0.9}
	\plotone{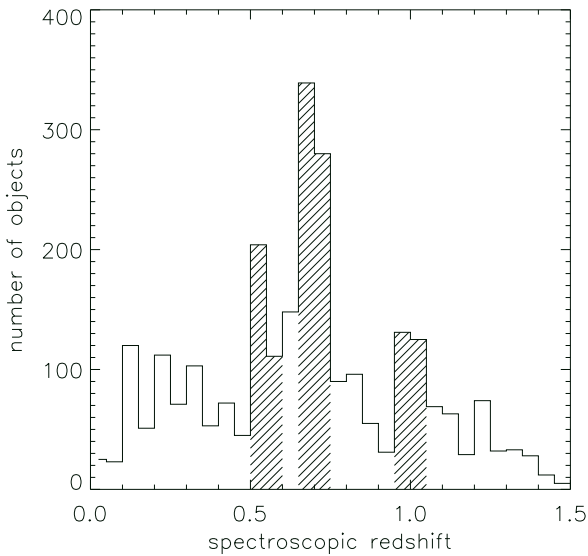}
	\vspace{0.3cm}
	\caption{The distribution of the spectroscopic redshift samples for the intermediate-redshift range. We used galaxies with redshifts around 0.55, 0.7 and 1.0 as indicated by the shaded areas, in order to minimize the contribution of $k$-correction uncertainties to the CMR scatter.}
	\label{redshift_histogram} 
\end{figure}

For the intermediate redshift range, up to $z=1$, we use data taken in the Extended Chandra 
Deep Field South (E-CDFS). 
A key ingredient is color HST imaging, taken from the 
\gems{}\footnote{Galaxy Evolution from Morphology and SEDs \citep{Rix}} and \goods{}\footnote{Great Observatories Origins Deep Survey \citep{Giavalisco}} surveys. These data were used to estimate accurate galaxy colors 
within the half-light radius, and for selection by galaxy structure.  
A second key ingredient, required to make
precise $k$-corrections, is accurate redshift measurements.  
For this investigation, a sample of objects with reliable spectroscopic redshifts was collected from a variety of sources.  Insecure or uncertain 
spectroscopic redshifts were cross-checked with photometric 
redshifts from the \combo{}\footnote{Classifying Objects by Medium-Band Observations in 17 Filters \citep{Wolf03,Wolf04}} survey. 
The spectroscopic selection criteria are described in the appendix and 
result in a final sample of 3440 galaxies with spectroscopic redshifts.  
We used 3030 galaxies in what follows; the remaining 410 objects could not be used either because they 
lacked data in one or more of the HST bands (211 galaxies), or 
because the object had a half-light radius smaller than 2 
pixels (204 objects and 5 objects satisfied both criteria). Figure \ref{redshift_histogram} shows the redshift distribution of the 3030 useable objects. We choose the three best populated redshift bins of width $\Delta z=0.1$ at $z \sim$ 0.55, 0.7 and 1.0 for our analysis. We made this selection for two reasons. Firstly, the bins have to be quite narrow to minimize the contribution of $k$-correction (the wavelength range over which the interpolation must be done) uncertainties to the final error budget. Second, the best populated redshifts had to be used in order to obtain sufficiently populated CMDs for deriving CMR properties. In Figure \ref{redshift_histogram} these redshift intervals are indicated by the shaded areas. 

For further analysis, we rejected known IR and X-ray sources\footnote{Leaving in IR and X-ray sources gives similar results; we conservatively remove such sources in order to minimize the contribution of young stars and non-stellar light to the CMR scatter.}. IR sources were identified by comparison with MIPS 24$\mu$m observations of the CDFS
from Spitzer \citep{Papovich04}. We use the 24{\micron} band owing to its high sensitivity
to obscured star formation and AGN activity.  The 80\% completeness
limit is 83$\mu$Jy, corresponding to approximate obscured SFR limits
of (5, 10, 17) $M_{\sun} yr^{-1}$ at redshifts of 0.55, 0.7 and 1.0 
respectively \citep{Bell07} using a \citet{Kroupa01} IMF. 
 We further excluded galaxies detected
in deep Chandra imaging of the CDFS.  The coverage is non-uniform, 
with an exposure time of 1Ms in the central pointing, and 250ks per 
pointing in each of 4 flanking fields.  These depths are 
sufficient to detect 
moderate-luminosity AGNs ($L_{0.5-2keV} = 10^{41} - 10^{42} \ ergs \ s^{-1}$)
over the whole redshift range of interest \citep{Lehmer05}.

\subsubsection{Colors and magnitudes from GEMS Images}

\begin{figure}[t]
	\centering
	\epsscale{0.9}
	\plotone{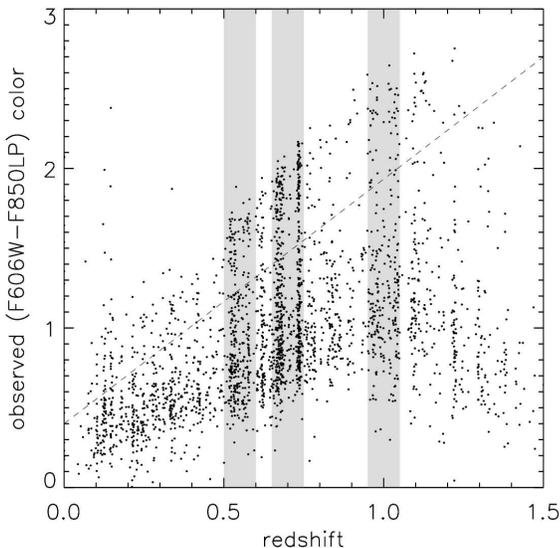}
	\vspace{0.3cm}
	\caption[Measured colors in $606-850$ as a function of redshift]{The observed F606W$-$F850LP color within the half light radius plotted as a function of redshift. The red sequence and blue cloud are clearly visible. The cut between the two populations is schematically indicated by the dashed line (this cut is only for illustration purposes and not used in what follows for any analysis; the actual redshift and magnitude-dependent cuts are given in Sec. 4.1). The gray shaded areas show the three redshift ranges used in this paper.}
	\label{vz_redshift}
\end{figure}

As the goal of this paper was to measure the scatter of the CMR as accurately as possible, we use high accuracy HST colors for the construction of the CMR.
For all objects with spectroscopic redshifts we used \gems{} postage stamps to measure accurate magnitudes in the F606W and F850LP passbands. We used the GALAPAGOS\footnote{Galaxy Analysis over Large Areas: Parameter Assessment by \galfit{}ting Objects from SExtractor (M.\ Barden et al.\ in preparation)} software package to cut postage stamps around the position of each object. Galaxy properties were adopted from a single-component
S\'ersic model \citep{Sersic} fit to the 2-D galaxy luminosity profile using the package \galfit{} \citep{Peng}. 
We estimate colors within the observed half-light radius (not the intrinsic half-light radius $r_{e,GALFIT}$ returned from fits to the light profile); such observed half-light radii are substantially larger than the 
intrinsic half-light radii for compact galaxies.  We determined the observed
half-light radius by performing aperture photometry in ellipses with the 
position angle and axis ratio given by \galfit{}, out to 
$10\,r_{e,GALFIT}$ on the F850LP-band image.  The {\it total} magnitude 
was defined as the aperture magnitude within $10\,r_{e,GALFIT}$. The observed
half-light radius
was then defined to be the semi-major axis of the ellipse 
within which half of the total light in F850LP was contained.  In order to determine 
accurate F606W$-$F850LP colors, the F606W-image was convolved with a difference
PSF, determined from the PSFs \citep{Jahnke} in F606W and F850LP (the PSF correction 
was accurate to within 1 part in a million in terms of total flux on a pixel-by-pixel basis; 
see \citealp{BorisThesis} section 4.2.2 for details), and the flux in F606W within the 
F850LP half-light ellipse was measured.
Figure \ref{vz_redshift} shows the distribution of the measured 
colors against redshift.  Uncertainties in the measured fluxes
include contributions from Poisson uncertainty, read noise, and 
uncertainty from inaccuracies in the assumed sky level (this
contribution is equal to the half-light area in pixels times the sky level
uncertainty in counts per pixel). Typical values for uncertainties in the determined magnitudes are around 0.007 (slightly higher for high redshifts and lower for smaller redshifts).

\subsection{The low-redshift sample from the Sloan Digital Sky Survey}

In order to explore the scatter of the CMR at low redshift, 
we use a sample of galaxies drawn from the 
Sloan Digital Sky Survey (SDSS) Data Release 4 \citep{SDSSDR4}. 
We choose a sample of galaxies from the 
publicly-available New York University Value-Added 
Galaxy Catalog \citep[NYU VAGC;][]{Blanton05nyuvagc} in a 
very narrow redshift range $0.0495 < z < 0.0505$ (2053 galaxies).
To maximize the accuracy of the color information, we adopt colors derived from
`Model' magnitudes --- such magnitudes
use the best-fit de Vaucouleurs or exponential models
in the $r$-band as a kernel for measuring fluxes in $ugiz$\footnote{We 
also used aperture magnitudes with aperture radii $5<r/{\rm arcsec}<10$, 
finding similar or larger CMR scatter.}.  We $k$-correct the observed-frame
$ugriz$ model colors to rest-frame $U-V$ color using InterRest (the same method we use for the intermediate redshift data; see 
Section \ref{sec_kcorr}). Rest-frame magnitudes were calculated by $k$-correcting S\'ersic model magnitudes \citep{Blanton} from the NYU VAGC; such magnitudes are closer to the total magnitudes than Petrosian or Model magnitudes. The formal random error in the final $U-V$ color
is $<0.02$\,mag from photometric error, with an estimated calibration
error of $<0.04$\,mag (assuming 0.01\,mag calibration errors in 
$g$ and $r$ bands); random errors in $M_{V}$ are $\sim 0.1$\,mag, dominated by systematic errors in how one defines sky levels and total magnitude.

\section{$k$-corrections}
\label{sec_kcorr}

\begin{figure}[t]
	\centering
	\epsscale{0.9}
	\plotone{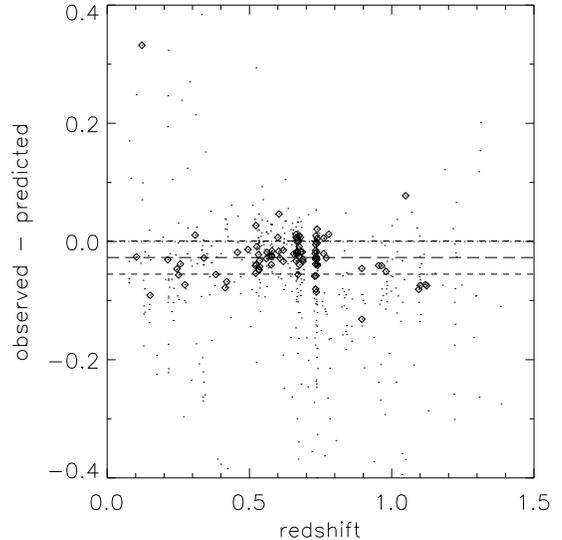}
	\vspace{0.3cm}
	\caption{An assessment of the $k$-correction uncertainties through comparison of the observed F775W magnitude of those sample galaxies that overlap with the \goods{} field and the F775W magnitude predicted using InterRest using only the redshift, F606W and F850LP magnitudes as input. In this plot we show the difference between measured and calculated magnitude in F775W as a function of redshift. The diamond symbols indicate red objects with a S\'ersic index $n > 2.5$. The outlier at $z \sim 0.1$ is a composite merging system with both a dust-reddened and a blue component; such a composite system is too complicated to be described by the simple templates used here. The long dashed line shows the mean of these values and the short dashed lines show the RMS area around the mean value (both values are around 0.03). The random error serves as an approximation of the $k$-correction uncertainty.}
	\label{kcorr} 
\end{figure}

To compare measurements in a redshift-independent manner, we $k$-correct the observed frame measurements into restframe properties (throughout this paper we use $M_{V,rest}$ and $(U-V)_{rest}$ although our conclusions do not depend on this choice). To do so we use the IDL implemented restframe interpolation code InterRest by ENT (http://www.strw.leidenuniv.nl/$\sim$ent/InterRest; this work is based on an earlier version of the code).  
To derive the redshift-dependent transformation between observed and rest-frame colors (the algorithm is described in more detail in Appendix C of \citealp{Rudnick03}), InterRest uses observed SEDs from a set of template galaxies: four empirical model spectra from \citet{Coleman80}, and one additional starburst template from \citet{Kinney96}; this is a subset of the default template set which is used to avoid degeneracies in color space.

These $k$-corrections were tested in two ways.  Firstly, one can use the $k$-correction routine to predict a F775W magnitude for galaxies in \goods{} (where one has F606W, F775W and F850LP).  We show such a comparison in Fig.\ \ref{kcorr}, 
where one can see that the $k$-corrections give a scatter of $\sim 0.03$ and an equal amount of mean color offset. Secondly, we can compare the final rest-frame colors to the results of independent $k$-correction codes.  Comparison to (lower accuracy) COMBO-17 colors
shows a scatter of 0.12 magnitudes and an offset of 0.06\,mag, and comparison to a stellar population model-derived $k$-correction by EFB and BH shows measurement-to-measurement scatter of less than 0.02 mag but overall rest-frame color 
offsets of $\sim0.1$\,mag (this results primarily from a difference between the 4000{\AA} break structure of the stellar population models and the observed ones used by InterRest).  We conclude that the $k$-corrections of the HST-derived colors are accurate to a few hundredths of a magnitude (random error; this contributes to the scatter on the CMR; Taylor et al., 2009, submitted), with possible overall systematics of $\lesssim 0.1$\,mag (affecting primarily the zero-point of the CMR).
We assume the $k$-correction errors to scale with the wavelength range over which the interpolation is done.

\section{Results}

In this section we present our analysis of the red sequence CMR scatter and its evolution. First, we explain the selection criteria applied for red sequence galaxies, then we discuss the actual measurement of the scatter of the red sequence at different redshifts 0.05 $\leq z \leq$ 1.05. With our precise magnitudes and colors we have produced color-magnitude diagrams (CMDs) for four redshift intervals and plot them in Figure \ref{cmd_uv}. 

\subsection{Selection of Red Sequence Galaxies}

Our selection criteria take care to choose only non-star-forming, early-type galaxies representing the red sequence and to reject blue galaxies, galaxies with disk structure, and galaxies with ongoing star formation and/or AGN activity.

\begin{figure*}[!htb]
	\centering
	\epsscale{2.1}
	\plotone{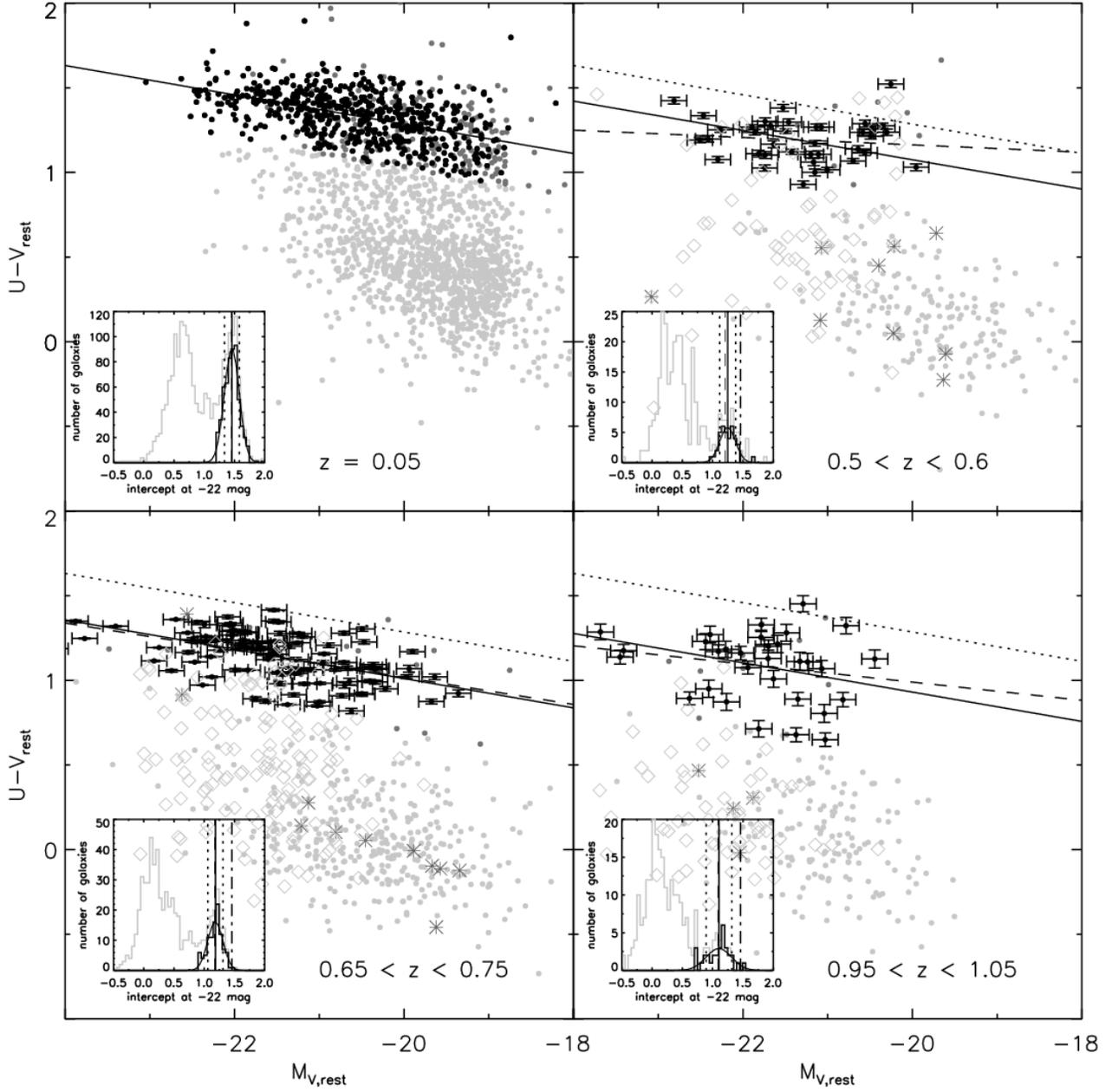}
	\caption[Color-magnitude diagram for $U-V$ against $M_{V}$ at restframe]{$U-V$ restframe color against $M_{V}$ for four different redshift bins. The light gray points are objects below a tilted cut, dividing blue and red objects. This cut has a fixed slope and a redshift-dependent intercept. Depending on the S\'ersic index the red objects are divided in a dark gray ($n<2.5$) and a black sample ($n>2.5$; with errorbars for the three intermediate redshift bins); the $n>2.5$ red galaxies are retained in our analysis. Independent of color and S\'ersic index IR (diamonds) and X-ray (asterisks) detections are rejected from the red sequence sample. The dashed lines are fits to the red sequence samples for the three redshift bins with $z\geq0.5$. In the $z=0.05$ panel the fit is shown as a solid line (the position of this fit is shown in the other panels as the dotted line). A line with the same slope is shifted to the mean color value of the three other panels (solid line). Using these `fits' we calculated intercepts at $M_V = -22$. The small histograms show the color distribution measured as an offset from the relation. The complete distribution is shown in light gray, whereas the red sequence (the black points in the CMDs) is shown in black. The dashed vertical line shows the intercept of the fit (dashed line in the CMDs) at a magnitude value of $-22$. The solid vertical line shows the mean intercepts with the CMR scatter (RMS) marked by the dotted vertical lines. The dash-dotted lines show the CMR intercept at $z=0.05$. Overplotted is a Gaussian with mean and $\sigma$ defined by the mean color and RMS of the red sequence sample.}
	\label{cmd_uv}
\end{figure*}

First, we applied a color cut to exclude galaxies with obvious signs of star formation from the sample. For this we made use of the bimodality of the color distribution which is visible in the CMDs at all redshifts (Figure \ref{cmd_uv}). The color cut was chosen to fall into the gap between blue and red objects (see also histograms in Fig. \ref{cmd_uv}). We used a tilted cut with the same slope for all samples. As the mean colors of the populations evolve with redshift we take this into account by allowing the cut to evolve correspondingly. The cut applied here can be described as $(U-V)_{rest} > -0.085 \cdot M_{V,rest}-0.65-0.5 \cdot z$. Small changes in the positions of the cuts do not influence the general results. In the diagrams the cut is made visible by different shades of gray. The light gray points are objects bluer then the color criterion whereas the dark gray and black points indicate object on the red side of the cut.

A second selection criterion was applied to clean the sample of structurally late-type galaxies. For this the S\'ersic index of the galaxy is used (for the \gems{} galaxies, this is measured using \galfit{}, while for the SDSS galaxies we adopt the S\'ersic fits from \citealp{Blanton05nyuvagc}). To be treated as part of the red sequence a S\'ersic index of $n>2.5$ is required. This cut successfully weeds out edge-on spirals from the sample \citep{McIntosh2005}. These objects are shown as dark gray points. Regardless of color and S\'ersic index we reject galaxies with X-ray or 24$\mu$m detections (for the CDFS data only; the SDSS sample
we use lacks such information) as their colors may be substantially affected by UV-bright young stellar populations and/or an accreting supermassive black hole (although our results remain unchanged if these systems are not excluded from our analysis). These objects are shown in the CMDs with asterisks (X-ray) and diamonds (IR). The objects which qualified as red sequence galaxies after these selection criteria are shown in black with error bars in the CMDs (number of RS galaxies in the redshift bins: 563/38/102/31).

\subsection{Fitting the Red Sequence}

The goal of this paper is to measure the intercept of the CMR at a fiducial magnitude and its scatter, and to show the evolution of both the CMR intercept and scatter as a function of redshift (see Figure \ref{cmd_uv}). The intercept is the color value of the relation measured at a specific magnitude (either the same magnitude value at all redshifts, or using a variable magnitude that attempts to account for the evolution of the population). The scatter is calculated using the IDL routine \textit{robust\_sigma}, which gives an outlier-resistant estimate of the dispersion of the color distribution using the median absolute deviation as an initial estimate and then weights points using Tukey's Biweight as an robust estimator \citep[this is the usual approach employed in this field; e.g.,][]{BLE92,McIntosh2005cluster}. We fit a simple linear function to the distribution of colors and magnitudes of the non-star-forming early-type
galaxies under two assumptions: 
\begin{enumerate}
	\item We fit the data points in each redshift bin separately.
          \item We choose to hold the slope fixed at all redshifts $\ge 0.5$ to the value measured at $z=0.05$ from the SDSS.
\end{enumerate}
The slope of the two methods are in some cases very different, but as the SDSS sample has many more objects and a larger dynamic range, this fit is much more reliable than the fits for the other bins. Further support for this assumed slope is given by the best populated \gems{} bin around $z=0.7$, which shows nearly the same slope. We note that the determination of the intercept and scatter are not sensitive to our choice of fitting method (Table \ref{table})\footnote{Furthermore the effect of excluding IR and X-ray detections from the red sequence is quite small and does not affect our results. The effect on the intercepts is around 1\%. The values for intrinsic scatter would be a bit larger if IR and X-ray sources are included in the sample, changing for $z=0.55$ from 0.128 to 0.138, for $z=0.7$ from 0.126 to 0.133 and for $z=1.0$ from 0.207 to 0.216.}. The histogram of color offsets from the CMR with redshift-independent slope is shown in the inset panels of Fig.\ \ref{cmd_uv} for early-type galaxies (black) and for all galaxies (gray). As the observed scatter contains contributions from intrinsic scatter and measurement uncertainty we estimate the intrinsic color scatter by subtracting the random color uncertainties in quadrature. The color uncertainties for the low redshift galaxies were estimated by scaling the Model $u-r$ color uncertainties\footnote{Recall that $U-V$ is estimated from the $ugriz$ Model colors using InterRest. Our choice of scaling from $u-r$ uncertainty is reasonable: $u$, $g$ and $r$ are the main determinats of $U-V$, where $g$ and $r$ are relatively well-measured and the $u$ band uncertainty is factors of several larger.} by $d(U-V)_{rest}/d(u-r)$, and adding in quadrature a small empirically-determined 0.015 mag contribution accounting for real scatter in $U-V$ at a given $u-r$: $\delta(U-V)^{2}=\left[ \delta(u-r) \cdot d(U-V)/d(u-r)  \right]^2 + 0.015^2 $. If not further specified we refer to the intrinsic scatter hereafter.

\begin{table*}
\caption{}
\begin{small}
\begin{tabular}{|c|cccc|cccc|}	
	
	\hline
	&& Fit &&&& Fixed slope &&\\
	&&& measured & intrinsic &&& measured & intrinsic\\
	& slope & intercept & scatter & scatter & slope & intercept & scatter & scatter\\

\hline
 $z=0.05$ & $-0.087$ $\pm$ 0.006 & 1.457 $\pm$ 0.005 & 0.124 & 0.117 & $-0.087$ & 1.457 $\pm$ 0.005 & 0.124
 & 0.117 \\ 
  $z=0.55$ & $-0.015$ $\pm$ 0.028 & 1.206 $\pm$ 0.021 & 0.126 & 0.124 &  & 1.248 $\pm$ 0.022 & 
0.130 & 0.129 \\ 
  $z=0.70$ & $-0.066$ $\pm$ 0.014 & 1.178 $\pm$ 0.010 & 0.127 & 0.127 &  & 1.184 $\pm$ 0.011 & 
0.126 & 0.126 \\ 
  $z=1.00$ & $0.010$ $\pm$ 0.047 & 1.097 $\pm$ 0.041 & 0.210 & 0.206 &  & 1.103 $\pm$ 0.036 & 
0.212 & 0.208 \\ 
\hline

\end{tabular} 
\end{small}
	\centering
\\[-5ex]
	\tablecomments{Color intercepts (at $M_{V,rest}=-22$) and scatter in $(U-V)_{rest}$ measured with two different methods. `Fixed slope' means that the slope of the CMR is fixed to the $z=0.05$ value for all redshifts (solid lines in Fig. \ref{cmd_uv}), while in the `Fit' columns the red sequence is fitted for each bin (dashed lines in Fig. \ref{cmd_uv}).}
	\label{table} 
\end{table*}

\begin{figure}[t]
	\centering
	\epsscale{0.9}
	\plotone{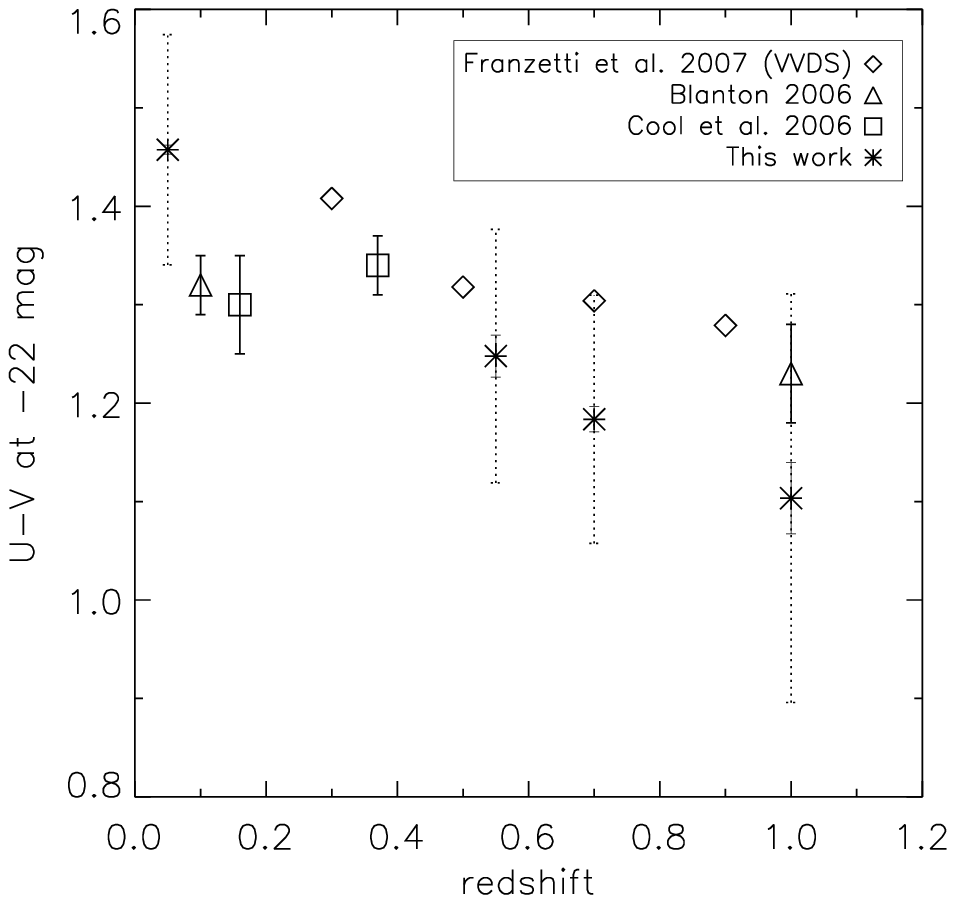}
	\vspace{0.3cm}
	\caption{CMR intercepts in $U-V$ as a function of redshift. The small solid errorbars denote the standard deviation of all intercept values calculated with the bootstrap method. The `errorbars' with dotted lines indicate the resistant dispersion of the color offsets from the best-fit CMR (i.e., the scatter
of the CMR). The other symbols show measurements of the CMR intercepts from other papers, as explained in the legend. The measurements from Blanton and Cool were transformed into $U-V_{rest}$ using kcorrect. The errorbars reflect the uncertainties introduced by this treatment.}
	\label{intercept_uv}
\end{figure}

The color intercepts $(U-V)_{rest}$ of the fitted CMR at a fixed $M_{V,rest}=-22$ are shown as a function of redshift in Figure \ref{intercept_uv}. The solid errorbars denote the formal error in the mean intercept, calculated using bootstrapping, although recall that the actual, primarily systematic, uncertainty in the galaxy colors is $\sim 0.1$\,mag.
The dotted `errorbars' show our measurement of the CMR scatter, calculated as the resistant dispersion of the color offsets of the early-type galaxies from the CMR. The results are tabulated in Table \ref{table}.
For comparison, intercept measurements from other studies \citep{Blanton06,Franzetti06,Cool06} are shown in this plot together with our data.  In the 
case of \citet{Blanton06} and \citet{Cool06}, we have $k$-corrected their 
values to rest-frame $U-V$ color using the $kcorrect$ software package, written by M.\ Blanton. We made sure that there is no significant difference ($\lesssim 0.03$\,mag) between the rest-frame colors derived with kcorrect and InterRest.

We find that the intercept of the color-magnitude relation evolves by $\sim 0.3$ mag in $U-V$ color between $z=1$ and $z=0.05$, in rough agreement with previous studies \citep[e.g.][]{Bell04a,Franzetti06,Blanton06,taylor2008}, although recall that systematic errors in the intercepts of the CMR are significant, $\sim 0.1$ mag in $U-V$ rest-frame color. More importantly, we find a scatter of $\sigma_{U-V} \sim 0.1$ mag (Table \ref{table}) at all $z < 0.75$; the scatter at $z=1$ appears to be somewhat larger \citep[being consistent with][]{taylor2008}. These values are somewhat larger than the scatter measured for local galaxy clusters ($0.04 < \sigma_{U-V} < 0.1$, where the scatter appears to vary from cluster to cluster; e.g. \citealp{McIntosh2005cluster})\footnote{The 
possible trend towards a bluer CMR intercept at very large clustercentric
radii, arguably an environmental effect, would also 
increase the scatter of the cosmic average CMR relative to 
the cluster CMR \citep[e.g.,][]{Terlevich01,Pimbblet02}.}. 
We note that subtraction in quadrature of the individual galaxy measurement
errors from the CMR scatter yields almost unchanged results, even 
for the low redshift SDSS sample (where the scatter could decrease to 
$\sim 0.09$\,mag). This shows that we measured a real scatter in the relation and not only a spread caused by measurement uncertainties. In particular, one should note that for the 
SDSS, spectral analysis of the drivers of the CMR scatter have demonstrated
clear spectral differences between early-type galaxies of a given magnitude 
at the red side and blue side of the CMR \citep{Gallazzi06,Cool06}, demonstrating that the bulk of the CMR scatter is intrinsic.

\section{Interpretation}

In the last sections, we described our measurements of the CMR intercept and scatter for cosmic average early-type galaxies at four
redshifts between 1.0 and 0.05.  We find a slowly-evolving mean 
color, and an almost non-evolving scatter of $\sim 0.1$\,mag in 
U-V rest-frame color. In this section, we build some intuition about possible interpretation of this result using stellar population synthesis modeling. We show first the evolution of single bursts, or populations of galaxies whose star formation is truncated at a particular time, in \S 5.1 and 5.2. Note that the purpose here is not to test `monolithic collapse' models (those models are already ruled out by the observed build up of the red sequence population), rather it is to establish the basic model ingredients. In \S 5.3 and 5.4 we ask and answer a simple question \citep[similar in spirit to][]{vanDokkumFranx01}: if the red sequence forms through ongoing and continious truncation of star formation in previously star-forming systems, can one simultaneously reproduce the build-up of red sequence galaxies and the scatter of the CMR?

\subsection{The evolution of single bursts}
\label{pass_ev_sec}

\begin{figure*}[!htb]
	\centering
	\epsscale{2.0}
	\plotone{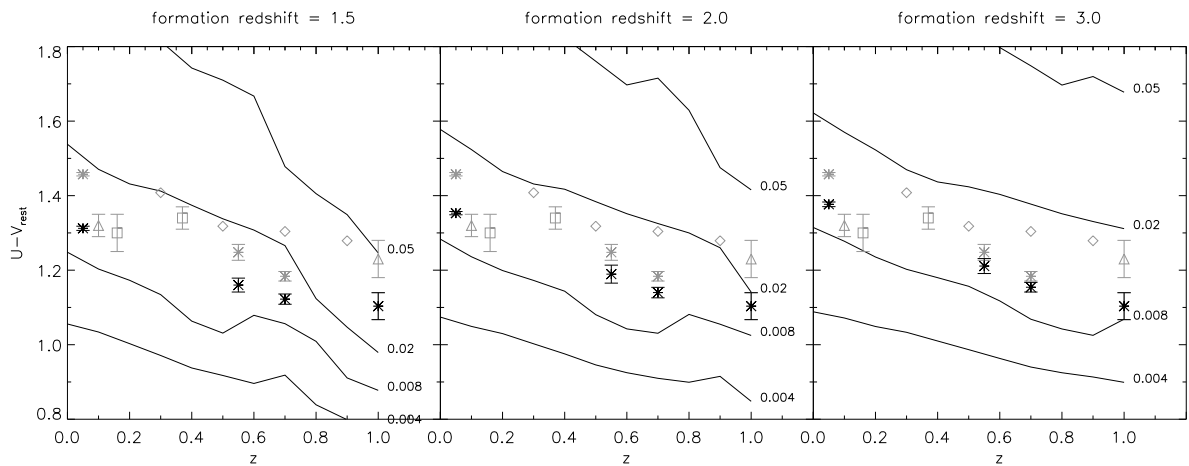}
	\vspace{0.3cm}
	\caption[Evolution with redshift for different metallicities]{The 
evolution in rest-frame color of single-burst stellar populations formed at three
different redshifts, and with different metallicities (where $Z = 0.02$ is 
solar metallicity).  The asterisks denote the measured intercepts
from our paper; the gray asterisks are the same as in Fig. \ref{intercept_uv}, whereas the black asterisks show color intercepts measured at a redshift-dependent absolute magnitude (to compensate for 
fading of stellar populations as the galaxy population ages; one can 
think of this as measuring the intercept at approximately constant 
stellar mass).  The other gray symbols show data from other papers, 
measured in a similar fashion to our gray data points (see the legends of Fig. \ref{intercept_uv} or \ref{trunc}).}
	\label{models}
\end{figure*}

To get our bearings, we focus first on the evolution of the CMR intercept using 
single bursts of star formation.  In what follows, 
we use the stellar population model 
\pegase{}\footnote{Projet d'Etude des GAlaxies par Synth\`ese Evolutive}
 \citep[Version 2.0, see][for the description of an earlier version of the model]{pegase}
to predict the evolution of galaxy colors and absolute magnitudes
as a function of redshift.  

The results are shown in Figure \ref{models} for three different formation redshifts and for various metallicities.  One can see that the evolution of
the intercept of the CMR follows roughly the trends expected for the 
passive reddening and fading of ancient stellar populations.  
In the context of single bursts, the rate of change
of the CMR intercept with redshift is related primarily to formation history, 
whereas the overall intercept is sensitive primarily to metallicity, with 
some sensitivity to age.  

To compensate for the passive fading of the early-type population, we also measure the CMR intercept at a {\it redshift-dependent} absolute magnitude (black asterisks in Figure \ref{models}). We 
choose to adopt a model with metallicity $Z=0.008$ and 
a formation redshift of 2 to estimate the evolution of the 
luminosity of an early-type galaxy as a function of redshift\footnote{The choice between the formation redshifts in the range shown in the plot does not have a great influence on the results; to choose $z_{f}=3$ instead would lead to in a slightly smaller evolution in magnitudes, but the resulting difference in intercepts is smaller than 0.02 for all redshifts. A different choice in metallicity has also only very small effects. To choose $Z=0.02$ (solar metallicity) would change the results even less than the change of formation redshift.}, and
use this to calculate the absolute magnitude values at which 
we measure the intercept of the CMR (these absolute magnitudes
are $-22$, $-21.5$, $-21.3$, and $-20.8$ at $z=1.0$, 0.75, 0.55 and 0.05 respectively).
One can think of these intercepts as being measured at 
a given, redshift-independent, stellar mass.  Because these
intercepts reflect the evolution at a given stellar mass, we 
choose to focus on these values in what follows, as a more intuitive 
reflection of the likely evolution of the color of a given 
galaxy (or a mass/metallicity bin in the evolving galaxy population).

\subsection{The evolution of galaxies with truncated star formation}

\begin{figure}[t]
	\hspace*{-1cm}
	\centering
	\epsscale{0.9}
	\plotone{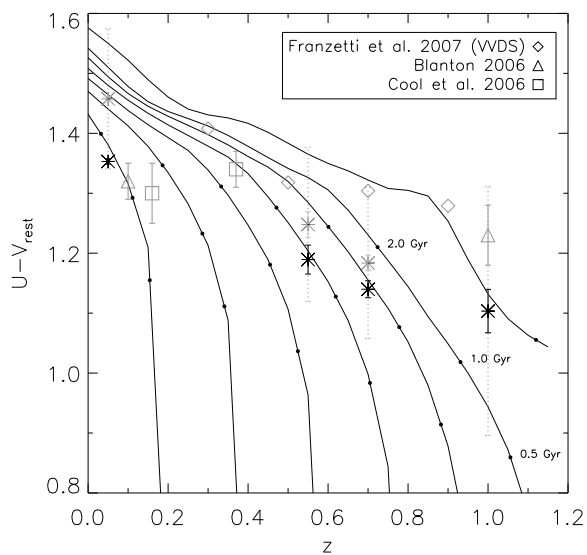}
	\vspace{0.3cm}
	\caption[Model tracks with truncated star formation.]{The redshift evolution of rest-frame color for different truncation times. For all lines, star formation starts at $z=2$ and the different lines denote a variety of different truncation redshifts (from left to right: 0.2 to 1.2 in steps of 0.2). The tiny black dots on the lines indicate the point on the evolutionary track a galaxy has reached 0.5, 1 and 2 Gyr, respectively, after the truncation of star formation (as labeled on the second line from the right). The symbols are the same as in Figures \ref{intercept_uv} and \ref{models}.}
	\label{trunc} 
\end{figure}

Motivated by the observational evidence for truncation of star formation in blue sequence galaxies, leading to the ongoing build-up of the red sequence galaxies at redshifts $z<1$ \citep[e.g.][]{Bell04b,Borch,Bell07}, we study the evolution of the colors of galaxies whose star formation has been truncated at a variety of redshifts $z \la 1$. The purpose of this section is to get a feeling for the ingredients used for the modeling of the CMR evolution described in the next section.
Figure \ref{trunc} shows the rest-frame color evolution of a stellar population which 
forms stars at a constant rate from $z_f = 2$ to the truncation redshift
$z_{\rm trunc}$. We plot the evolution of seven different truncation histories which all have a constant metallicity of $Z=0.02$. The rightmost track has a truncation redshift very close to the formation redshift of the model, namely at $z=1.95$. The other lines show the evolution of galaxies with smaller truncation redshifts (from left to right) from 0.2 to 1.2 in steps of 0.2. 
A general feature of such truncation models is a period of relatively
rapid evolution onto the red sequence (timescales $\lesssim 1$\,Gyr, as can be seen in Fig. \ref{trunc} and \citealp{Blanton06,SchweizerSeitzer92}), and then relatively slow
subsequent fading and reddening of the population.

\subsection{Expectations for the Scatter Measurement}

\begin{figure}[t]
	\centering
	\epsscale{0.9}
	\plotone{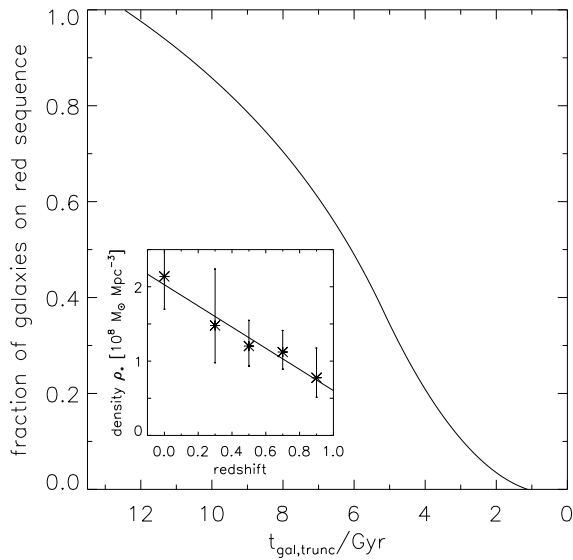}
	\vspace{0.3cm}
	\caption{The integrated number of truncated (pre-red sequence) galaxies as a function of cosmic age ($t_{gal,trunc}=13.5$ Gyr is the present day). Assuming that after truncation of star formation galaxies fade onto the red sequence in 1 Gyr, one reproduces (by construction) the observed evolution in stellar mass density on the red sequence (inset panel).}
	\label{RS_buildup}
\end{figure}

With these SF modeling ingredients in place, we are now in a position 
to ask what the evolution of the intercept and scatter of the CMR 
can tell us about the evolution of the early-type galaxy population.
Obviously, we will be unable to address this issue completely, as we have
seen that many factors influence the colors and magnitudes of 
galaxies: when star formation started, metallicity, when (if!) star
formation ends, and whether or not there is any residual star formation. 

Instead, 
we limit ourselves to one well-defined question.  
Recent observations have measured a significant increase
in the total stellar mass in the red sequence galaxy population 
since $z=1$ \citep{Chen,Bell04b,Bundy05,Brown2006,Borch,Faber07}.
This evolution manifests itself primarily in terms of an increasing
space density of red sequence galaxies at $\la L^{*}$ \citep{Borch,Faber07}, and is most naturally interpreted as being fed
by the truncation or quenching of star formation in massive blue galaxies
\citep{Bell04b,Bell07,Faber07}. In such a scenario, 
one expects a significant scatter in the CMR, because of the constant 
flow of recently star-forming galaxies onto the red sequence.  Here, 
we will predict the CMR scatter implied by a constantly growing 
red sequence, and compare this predicted scatter with the observations. Such an exercise was carried out by \citet{vanDokkumFranx01} for early-type galaxies in clusters; here, we extend their work to a cosmic average environment.

Our first ingredient is a toy model for the build up in the 
number of red sequence galaxies as a function of time.  We assume 
i) that the growth in the total mass on the red sequence is driven entirely by adding blue galaxies that have had their SF recently truncated, and ii) red sequence galaxies are added at all stellar masses equally\footnote{In fact the CMR scatter and mass build-up are both measured at $\sim 10^{11}M_\odot$, reducing the importance of this assumption.}. 
We describe the evolution of the number of red sequence 
galaxies with a simple linear fit (in redshift) as shown in the small panel of
Fig.\ \ref{RS_buildup}, compared with the measurements of the integrated stellar mass density of \citet{Borch}.
We then determine the derivative of this relation to get the change in stellar mass density with time. This is used to estimate the {\it truncation rate} by shifting the time axis by $-1$ Gyr - approximately the time taken to redden enough to satisfy our red sequence cut (following \citealp{Blanton06}; see also Fig. \ref{trunc}). In this way, we have a truncation rate history that is consistent with \citet{Borch} and can be used to estimate the evolution of the CMR scatter. To describe the truncation rate for epochs not covered
by \citet{Borch}, we choose to model the truncation rate as a constantly increasing 
function of time from $z_f \sim 5.5$ until $z_{trunc} = 0.9$. The influence of this choice on the results is negligible. The resulting number of truncated (pre-red
sequence) galaxies -- the integral of the truncation rate -- as a function of cosmic epoch is also given in Fig.\ 
\ref{RS_buildup}.  

\begin{figure*}[t]
	\centering
	\epsscale{2.0}
	\plotone{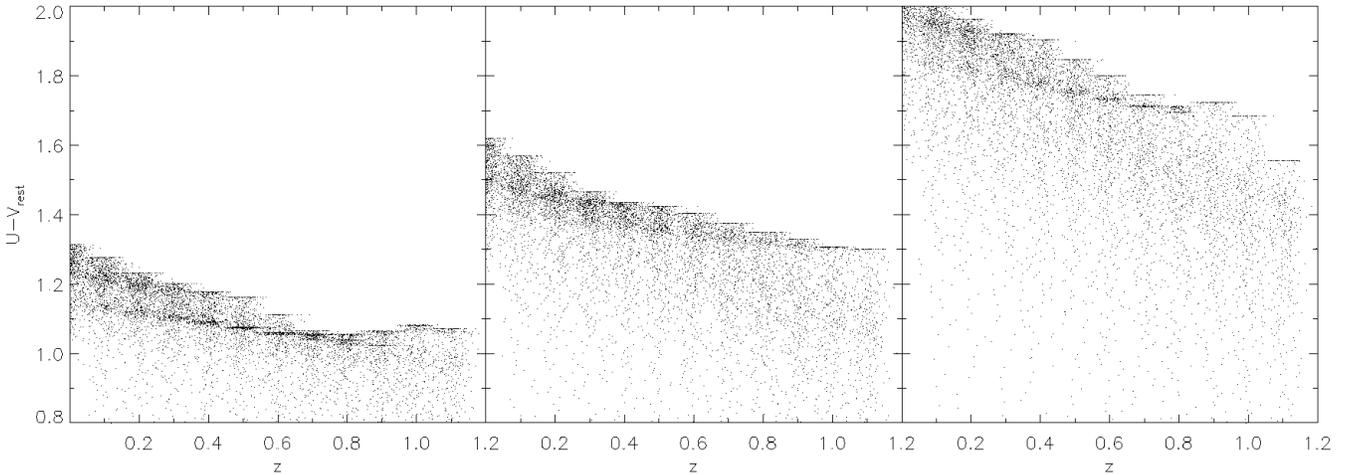}
	\vspace{0.3cm}
	\caption{Colors of individual galaxies calculated using the truncation history shown in Fig.\ \ref{RS_buildup} using one fixed metallicity for each plot. The colors were measured for certain redshift values and are shown here spread out in redshift to illustrate the density in color space. The metallicities are from left to right 0.008, 0.02 (solar) and 0.05. There is a bimodality at some redshifts, as a result of kinks in the evolutionary tracks at some values of age/metallicity.}
	\label{diff_met} 
\end{figure*}

\begin{figure}[t]
	\centering
	\epsscale{0.9}
	\plotone{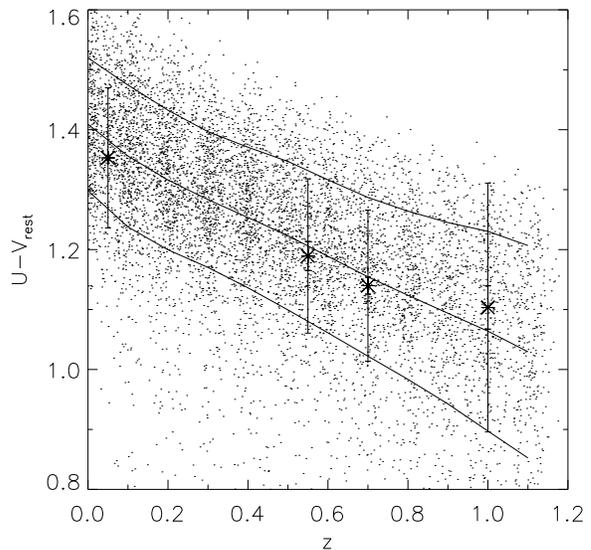}
	\vspace{0.3cm}
	\caption[Modeled red sequence evolution]{Colors of individual galaxies calculated in the same way as for Figure \ref{diff_met} but assuming a scatter in metallicity. The lines show the mean value in color and the scatter of the color distribution. Recall that we have used a redshift-dependent luminosity cut as described in Section \ref{pass_ev_sec} to measure the intercepts.}
	\label{final_model} 
\end{figure}

\begin{figure}[t]
	\centering
	\epsscale{0.9}
	\plotone{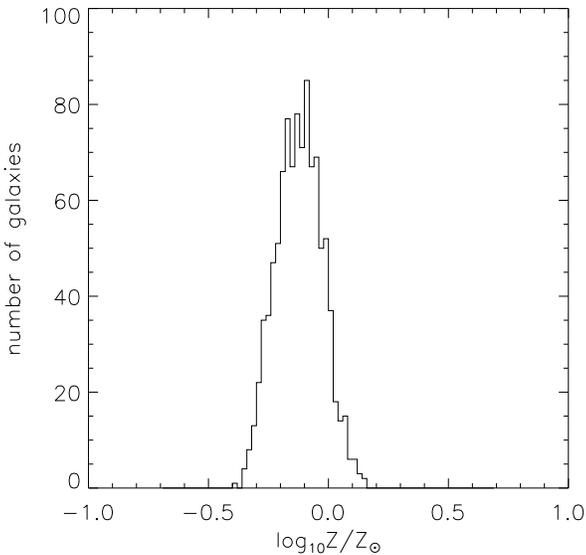}
	\vspace{0.3cm}
	\caption[Metallicity distribution for the modeled galaxy sample]{Metallicity distribution for the modeled galaxy sample in Figure \ref{final_model}. The mean value $\langle log_{10} Z/Z_{\odot} \rangle$ is $-0.12$ and standard deviation is 0.1. Solar metallicity is taken to be 0.02.}
	\label{met_distr} 
\end{figure}


\begin{table*}[!htb]
\caption{}
\begin{small}
\begin{tabular}{|c|cc|cc|c|}
	
	\hline
	& Fit && Fixed slope && Simulation\\
	& measured & intrinsic & measured & intrinsic &\\
	& scatter & scatter & scatter & scatter &\\

\hline
$z=0.05$ & 0.124 & 0.117 & 0.124 & 0.117 & 0.115 \\ 
$z=0.55$ & 0.126 & 0.124 & 0.130 & 0.129 & 0.126 \\ 
$z=0.70$ & 0.127 & 0.127 & 0.126 & 0.126 & 0.132 \\ 
$z=1.00$ & 0.210 & 0.206 & 0.212 & 0.208 & 0.167 \\ 
\hline

\end{tabular} 
\end{small}
	\centering
\\[-5ex]
\tablecomments{For better comparison the measurement values for the measured and intrinsic red sequence scatter are presented here together with the values for the simulated scatter.}
	\label{table_diss} 
\end{table*}

The next step is to build a more realistic truncation model (c.f. Fig. \ref{trunc}): instead of choosing one particular truncation time as in \S 5.2, we instead draw truncation times from the truncation time distribution in Fig. \ref{RS_buildup}. We choose 12 `observation' redshifts between $z=0$ and $z=1.1$; at each we draw 1000 galaxies from this truncation\footnote{The results of this paper do not change significantly if a burst of star formation occurs before truncation, as might be expected in a gas-rich galaxy merger.} history shown in Fig.\ \ref{RS_buildup} resulting in samples of galaxies with different truncation redshifts, but the same formation redshift $z_{start} = 3$ for all galaxies\footnote{Note that this choice of $z_{start}$ differs from $z_f$ in section \ref{pass_ev_sec}. In contrast to this former $z_f$, which was used to simulate the passive evolution after an initial star burst, $z_{start}$ indicates the starting point of a longer period of star formation.}. Due to this variety of truncation redshifts the galaxies in the samples experienced periods of passive evolution of different length. This then gives 
a distribution of galaxy colors at each redshift of interest; for display 
purposes in Figs.\ \ref{diff_met} and \ref{final_model} we have 
added a small random offset in redshift. We assume no ongoing low-level star formation in red sequence galaxies; while potentially unrealistic, it allows us to estimate the expected scatter from truncation alone\footnote{One could argue we will estimate lower limits to the scatter of the CMR with this approach. Such an argument has merit, but would neglect (as we also have) the influence of any age/metallicity anti-correlation on the CMR scatter \citep[as observed by e.g.][]{Trager2000b}}.

In order to separate between the effects of age and metallicity
scatter on the CMR scatter, we first consider the effects of formation 
history alone in Fig.\ \ref{diff_met}.  One can see a considerable color
scatter from the ongoing accretion of recently-truncated galaxies,
along with some low-level bimodalities caused by kinks in the time
evolution of colors for some metallicities (see, e.g., Fig.\ \ref{models}).
For solar metallicity $Z=0.02$ the color evolution is smooth, giving no 'pile-ups' in color space.  The strong influence of metallicity on the color of the red sequence can also be seen in this plot.

In Fig.\ \ref{final_model}, we show our expectation for the evolution 
of the CMR intercept and scatter as a function of redshift.
We have drawn galaxies from a log-normal distribution 
in metallicity with mean [Fe/H]$=-0.12$ and scatter $0.1$\,dex which is held constant over redshift (Fig. \ref{met_distr})\footnote{Our assumption of a redshift-independent scatter is clearly an over-simplification; yet, in the context of our model in which each galaxy has its own single metallicity it is a defensible one. In reality, the metallicity may evolve if there is any low level residual star formation in early-type galaxies (indeed, if there is an age-metallicity relation in the stars in an individual galaxy, its light-weighted age can evolve as the galaxy ages even in the absence of star formation).}. This 
distribution was chosen to approximately match the observed CMR intercept and 
scatter, but it is interesting to note that this value of 
metallicity scatter
is consistent with the estimated intrinsic scatter in metallicity of present-day early-type galaxies of 
$\sim 0.1$ dex from \citet{Gallazzi06}. We show the mean and scatter (calculated with the same estimation algorithm as the scatter in the observed relations) of the distribution as solid lines.  
The resulting distribution agrees well with the scatter measurements at the redshifts 0.05, 0.55 and 0.7. At $z=1.0$, the measured scatter exceeds the model 
prediction (allowing for more scatter in star formation histories).  
The results are presented in Table \ref{table_diss}.

It is interesting to compare our results with a similar analysis of 
early-type galaxies in clusters carried out by \citet{vanDokkumFranx01}. They use a similar 
model of a constantly-growing red sequence to predict both the 
mass-to-light ratio and color evolution (and scatter) of the early-type 
galaxy population, finding a relatively slow color evolution and a 
relatively constant scatter. Our results 
are in qualitative accord with their results 
for the cluster early-type galaxy population, but are representative of 
the cosmic-averaged early-type galaxy population and are constrained to 
reproduce the observed build-up of the early-type galaxy population.

\subsection{Blue Spheroids}

\begin{figure}
	\centering
	\epsscale{0.9}
	\plotone{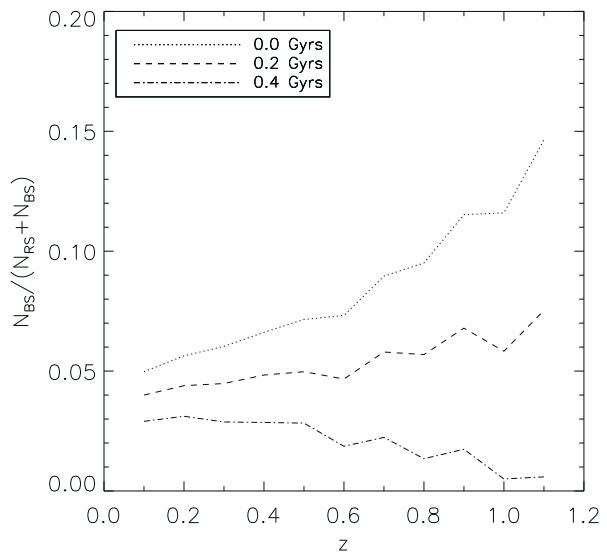}
	\vspace{0.3cm}
	\caption{The predicted ratio of blue spheroids to the total number of spheroids as a function of redshift for our toy model for the growth of the red sequence. The three curves show the effect of three different choices for $t_{recognize}$, the time taken for a galaxy to become recognizably early type after the event that truncates star formation. Despite the dependence of $N_{BS}/(N_{RS}+N_{BS})$ on $t_{recognize}$, it is remarkable that the predicted blue spheroid fraction is in the range of the observations \citep[e.g.][]{Haeussler07}.}
	\label{bluespher}
\end{figure}

So far, we have made no particular assumption on the mechanism by which 
galaxies quench their star formation and join the red sequence.  In this 
section, we explore a rather more specific scenario: galaxies 
structurally transform into an early-type galaxy (e.g., through galaxy 
merging) {\it before} reddening on to the red sequence.  Such a scenario 
predicts that there should be a non-negligible population of blue 
spheroids -- i.e., galaxies that are structurally early-type but have 
blue colors.  In this context, we present a range 
of predictions for their abundance, requiring that blue spheroids i) have colors at least 0.25 mag bluer in $U-V$ than the locus of red sequence galaxies at that redshift, and ii) are at least a given time $t_{recognize}$ from the truncation (or spheroid creation) event. This latter criterion is introduced to allow for a delay $t_{recognize}$ between the event that truncated star formation (e.g. a galaxy merger) and the galaxy becoming recognizably early type; in what follows we explore the range $0\leq t_{recognize} \leq 0.4$ Gyr, motivated by simulations of galaxy mergers. The ratio of blue spheroids to the total number of spheroidal galaxies (red and blue spheroids) $N_{BS}/(N_{RS}+N_{BS})$ is measured as a function of redshift and is shown in Fig. \ref{bluespher} for 3 different values of $t_{recognize}$.

There are two key points to take away from Fig. \ref{bluespher}. First, the predicted blue spheroid fraction depends sensitively on one's choice of $t_{recognize}$ (and therefore, also, details of the star formation history and dust content of the galaxy)\footnote{In fact, for long $t_{recognize}$, the blue spheroid fraction decreases towards higher redshift because $t_{recognize}$ becomes comparable to the time taken to transition between the blue cloud and red sequence.}. Second, despite this uncertainty, the range in model prediction is in reasonable agreement with observed blue spheroid fractions (e.g. $N_{BS}/(N_{RS}+N_{BS}) \sim 0.06$ at $z \sim 0.6$ from \citealp{BorisThesis}\footnote{Here we are comparing with the blue spheroid fraction of relatively 
massive spheroidal galaxies, which have a mass range consistent with our red sequence galaxies \citep[see][Fig. 4.13]{BorisThesis}. As \citealp{BorisThesis} (see paragraph 4.5.1) found many of the works quoting a higher 
blue spheroid fraction are including low-mass / low-density objects that 
can never turn into a present-day bulge-dominated red sequence galaxy \citep[e.g.,][]{Abraham1999,schade99,Im2002}.}). The fraction observed by \citet{Bamford2008} for the relevant mass range is slightly higher than in our models, but as they used a different method these values might not be directly comparable. Recall that this toy model was constructed to explore the implications of the growth of the red sequence through the truncation of star formation in blue galaxies. Here we have shown that such a model reproduces \textit{simultaneously} the evolution of the intercept and scatter of the color-magnitude relation since $z=1$ and the blue spheroid fraction at intermediate redshift. Such an analysis lends considerable weight to the notion that the early-type galaxy population has grown considerably between redshift one and the present day through the truncation of star formation in blue galaxies (through mergers or some other physical process).

\section{Discussion and Conclusions}

The evolution of the scatter of the red sequence is an important
source of information about the evolutionary history of 
the red sequence/early-type galaxy population.  
In this paper, we constructed high-accuracy color--magnitude relations
at four different redshift ranges using accurate color and spectroscopic 
redshift information.
We used a sample of over 3000 galaxies in the CDFS, in conjunction 
with a local reference sample of galaxies from the SDSS, to understand 
the redshift evolution of CMR scatter.
We used images from the \gems{} 
survey taken with the ACS onboard HST to 
provide both high-resolution morphologies
(to classify galaxies as early-type) and to measure accurate colors within 
the half-light radius.  In order to calculate accurate $k$-corrections to 
rest-frame passbands, we used spectroscopic redshifts compiled from a variety 
of sources.  At all redshifts, we apply a structure and color cut
to isolate early-type red-sequence galaxies; at intermediate redshift 
we were able also to excise star-forming galaxies and AGN from the sample
using X-ray and 24{\micron} information.

The resulting scatter of the color magnitude relation is $\sim 0.1$ mag
in $U-V$ color  
at $z=0.05$, 0.55 and 0.7, and somewhat higher at $z=1.0$.  
This scatter is comparable to those found in some local galaxy clusters (e.g. Abell 85 and 754 of \citealp{McIntosh2005cluster}; their Table 8) and larger than that found for the Coma and Abell 496 clusters ($\sigma_{U-V} \sim 0.05$ mag; \citealp{BLE92,Terlevich01,McIntosh2005cluster}). We note in passing that a 
better observational handle on the environmental dependence of the 
scatter of the color--magnitude relation, 
with special attention being paid to possible 
cluster-to-cluster differences, would be highly desirable (but 
is clearly beyond the scope of this paper).

We explored the implications of a measured CMR scatter of $\sigma_{U-V} \sim 0.1$ mag on our understanding
of the evolution of early-type galaxies. The CMR scatter can be influenced
by when star formation started, when star formation stopped, 
residual star formation and metallicity. Thus, it is not straightforward to 
turn the measured CMR scatter into robust constraints in this multi-dimensional space.  

We adopted instead a simple approach. The evolving number density
of red sequence galaxies has been measured --- assuming that this evolution 
in number density is from the quenching of star formation in galaxies with 
roughly constant star formation rate before quenching, we have a prediction 
(motivated by other observations) for the quenching/truncation history
of galaxies going on to the red sequence. We have also a measurement of 
the present-day metallicity scatter of early-type galaxies from \citet{Gallazzi06}.
Thus, we have both elements in place where we can {\it predict} 
the evolution of the CMR scatter. Importantly, this model --- 
in which new red sequence galaxies are
being constantly added at the rate required by observations of 
red sequence galaxy {\it number density} --- predicts the correct 
scatter in the {\it colors} of these red sequence galaxies.  

Furthermore, this model predicts approximately the correct number density of blue spheroids - galaxies which are structurally early-type but have blue colors - although admittedly with considerable model dependence. Thus, we conclude that these different observations - the evolution of the number density of red sequence galaxies, the evolution of the intercept and scatter of the color-magnitude relation, and the blue spheroid fraction - paint a consistent picture in which the early-type galaxy population grows significantly between $z=1$ and the present day through the quenching of star formation in blue cloud galaxies.

\acknowledgements

We thank the anonymous referee for constructive comments that improved the clarity of the paper significantly.
Based on observations taken with the NASA/ESA {\it Hubble Space
Telescope}, which is operated by the Association of Universities
for Research in Astronomy, Inc.\ (AURA) under NASA contract NAS5-26555.
Support for the GEMS project was provided by NASA through grant 
number GO-9500 from the Space Telescope Science Institute.
C.\ R.\ and E.\ F.\ B.\ acknowledge support
from the Deutsche Forschungsgemeinschaft through the Emmy Noether 
Programme.
M.\ B.\ was supported by the Austrian Science Foundation FWF under grant P18416.
D.\ H.\ M.\ acknowledges support from NASA under
LTSA Grant NAG5-13102.
Observations have been carried out using the Very Large Telescope at the ESO Paranal Observatory under Program ID(s): 170.A-0788, 074.A-0709, 171.A-3045 and 275.A-5060.

\newpage

\appendix
\section*{Spectroscopic Redshift Sample}
 
\begin{table*}
\caption{Spectroscopic redshifts from different catalogs}
\begin{small}
\begin{tabular}{|c|c|}

	\hline
	redshift catalog & number of objects used\\

\hline
VVDS \citep{LeFevre2004} & 629\\
\citet{Szokoly04} & 38\\
\goods{}/FORS2 DR 3.0 \citep{Vanzella08} & 464\\
\citet{Ravikumar06} & 303\\
S.\ Koposov et al., in prep. & 510\\
VIMOS DR 1.0 \citep{Popesso08} & 1086\\
\hline

\end{tabular} 
\end{small}
	\centering
	\label{table_redshifts} 
\end{table*}

\begin{figure}[t]
	\centering
	\epsscale{0.9}
	\plotone{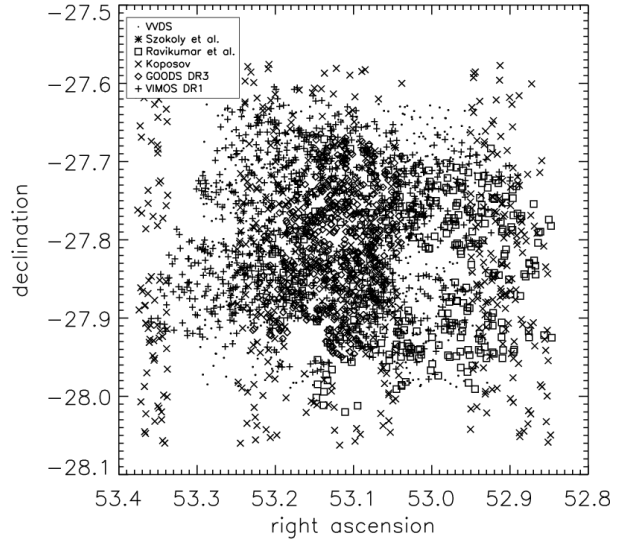}
	\vspace{0.3cm}
	\caption{All objects with spectroscopic redshifts adopted in this paper are shown here at their positions on the sky. The different symbols showing the origin of the spectroscopic redshift value for each object.}
	\label{objects} 
\end{figure}


In order to study the evolution of the scatter of the red sequence, one must use a sample of galaxies with spectroscopic redshifts. Toward this end, we compiled a catalog of spectroscopic redshifts from a variety of sources, listed below.

We used the data of some projects which surveyed the entire CDFS but a better coverage is reached for the \goods{} field (see also Fig. \ref{objects}). A collection of spectroscopic redshifts for the CDFS can be found in the `Master Catalog of CDFS spectroscopy' (http://www.eso.org/science/goods/spectroscopy/ \\CDFS\_Mastercat/). We used data from one paper included in this catalog \citep{Szokoly04}. Also the VVDS \citep[VIMOS-VLT Deep Survey;][]{LeFevre2004}, \citet{Ravikumar06} and S. Koposov et al. (in preparation; VVDS high resolution spectroscopy) surveyed the CDFS. With data limited to the \goods{} field we used the VLT/FORS2 Spectroscopy Data Release 3.0 \citep{Vanzella05,Vanzella06,Vanzella08} and the \goods{} VLT/VIMOS Spectroscopy Data Release 1.0 \citep{Popesso08}.
For sources with good quality flags the spectroscopic redshifts were taken without further testing. These are:

\begin{itemize}
\item VVDS - 4 (confidence level 100 per cent) and 3 (95 per cent)
\item FORS2 - A (solid redshift determination) and B (likely redshift determination)
\item \citet{Szokoly04} - 3.0 (reliable redshift determination with unambiguous X-ray counterpart) and 2.0 (reliable redshift determination)
\item \citet{Ravikumar06} - secure observations
\end{itemize}

In the case of insecure or 
unknown data quality, the spectroscopic
redshifts were tested against photometric redshifts from the \combo{} survey.  In such cases we used only those spectroscopic redshifts which differ by less than 0.1 from the photometric ones.
In total this leads to a sample of 3440 objects (3030 useable for this project). See Figure \ref{objects} for the positions on the sky of the objects used for flux measurements. 
The total numbers of objects of each catalog is listed in 
Table \ref{table_redshifts}. 
The combined sample covers a roughly constant fraction of about 20\% of the galaxies in the relevant magnitude range ($20\,mag< M_{I} < 22\,mag$) and then drops off towards fainter magnitudes with a fraction of about 10\% for $M_{I} < 24\,mag$. Red and blue galaxies show a similar behaviour.



\end{document}